\documentclass[traditabstract]{aa}

\usepackage{natbib}
\usepackage{graphicx}
\usepackage{txfonts}
\usepackage{mathptmx}

\begin{document}

\title{Deep \emph{Chandra} observations of the stripped galaxy group falling into Abell 2142}
\author{D. Eckert\inst{1,2} \and M. Gaspari\inst{3\thanks{Einstein and Spitzer Fellow}} \and M. S. Owers\inst{4,5} \and E. Roediger\inst{6} \and S. Molendi\inst{2} \and F. Gastaldello\inst{2} \and S. Paltani\inst{1}  \and S. Ettori\inst{7,8} \and T. Venturi\inst{9} \and M. Rossetti\inst{2} \and L. Rudnick\inst{10}}
\institute{
Department of Astronomy, University of Geneva, ch. d'Ecogia 16, 1290 Versoix, Switzerland\\
\email{Dominique.Eckert@unige.ch}
\and
INAF - IASF-Milano, Via E. Bassini 15, 20133 Milano, Italy
\and
Department of Astrophysical Sciences, Princeton University, 4 Ivy Lane, Princeton, NJ 08544-1001 USA
\and
Australian Astronomical Observatory, PO Box 915, North Ryde, NSW 1670, Australia
\and
Department of Physics and Astronomy, Macquarie University, NSW, 2109, Australia
\and
E.A. Milne Centre for Astrophysics, University of Hull, Hull, HU6 7RX, United Kingdom
\and
INAF - Osservatorio Astronomico di Bologna, Via Ranzani 1, 40127 Bologna, Italy
\and
INFN, Sezione di Bologna, viale Berti Pichat 6/2, 40127 Bologna, Italy
\and
INAF - Istituto di Radioastronomia, via Gobetti 101, 40129, Bologna, Italy
\and
Minnesota Institute for Astrophysics, School of Physics and Astronomy, University of Minnesota, 116 Church Street SE, Minneapolis, MN 55455, USA
}

\abstract{In the local Universe, the growth of massive galaxy clusters mainly operates through the continuous accretion of group-scale systems. The infalling group in Abell 2142 is the poster child of such an accreting group, and as such, it is an ideal target to study the astrophysical processes induced by structure formation. We present the results of a deep (200 ks) observation of this structure with \emph{Chandra} that highlights the complexity of this system in exquisite detail. In the core of the group, the spatial resolution of \emph{Chandra} reveals a leading edge and complex AGN-induced activity. The morphology of the stripped gas tail appears straight in the innermost 250 kpc, suggesting that magnetic draping efficiently shields the gas from its surroundings. However, beyond $\sim300$ kpc from the core, the tail flares and the morphology becomes strongly irregular, which could be explained by a breaking of the drape, for example, caused by turbulent motions. The power spectrum of surface-brightness fluctuations is relatively flat ($P_{2D}\propto k^{-2.3}$), which indicates that thermal conduction is strongly inhibited even beyond the region where magnetic draping is effective. The amplitude of density fluctuations in the tail is consistent with a mild level of turbulence with a Mach number $M_{3D}\sim0.1-0.25$. Overall, our results show that the processes leading to the thermalization and mixing of the infalling gas are slow and relatively inefficient.}

\maketitle

\section{Introduction}

Structure formation in the Universe operates as a bottom-up process in which small structures hierarchically merge to form larger systems. The imprint of structure formation is observable today through the continuous accretion of galaxies and groups onto massive clusters \citep{kravtsov12}. These processes are observable in the outer regions of local clusters, which are currently largely unexplored \citep{reiprich13}. Cluster outskirts host the transition region between the virialized intracluster medium (ICM) approximately in hydrostatic equilibrium, and the infalling material from the surroundings. 

Accreting galaxies and galaxy groups are commonly observed around massive galaxy clusters along preferential directions set by the large-scale structure \citep[e.g.,][]{einasto01,girardi02}. When falling onto a massive cluster, the gas content of the infalling structure interacts with the hot ICM and is stripped by the ram pressure applied by the surrounding gas \citep{gunn72,heinz03,takizawa05}, leading to the virialization of the accreting gas into the main halo. 

This picture is confirmed by observations of trails of stripped gas lagging behind infalling galaxies in X-rays \citep{sun05,machacek05,randall08,su17} and H$\alpha$ \citep[e.g.,][]{sun07,zhang13,boselli16}. Recently, X-ray observations have also revealed trails of stripped gas in merging galaxy groups \citep{neumann01,durret05,eckert14b,ichinohe15,degrandi16,sasaki16}. The accretion of groups with a mass on the order of a few $10^{13}M_\odot$ constitutes the main channel through which massive clusters accumulate mass and member galaxies at the current epoch \citep{dolag09,berrier09,genel10}. Thus, observations of such infalling groups are crucial for studying the growth of structures at late times and the processes through which the infalling material is virialized within the larger halo.

In addition, numerical simulations have shown that the properties of the stripped gas tails strongly depend on ill-constrained plasma parameters. The long survival of these trails implies that thermal conduction is strongly inhibited in the medium \citep{eckert14b,sanders14}. It is unclear, however, if thermal conduction is suppressed only at the interface between the two plasmas by magnetic draping \citep{asai05,ettori00}, or if conduction is inhibited throughout the plasma, for example, because of a tangled magnetic field configuration \citep{chandran98,ruszkowski10}. Additionally, \citet{roediger15a,roediger15b} used high-resolution numerical simulations to show that the morphology of the stripped tails depends on the viscosity of the plasma. In the inviscid case, Kelvin-Helmholtz instabilities rapidly develop and induce a substantial level of turbulence in the wake. Conversely, a high viscosity stabilizes the flow and creates long X-ray trails.

Abell 2142 is a massive galaxy cluster \citep[$M_{200}\sim1.3\times10^{15}M_\odot$,][]{munari14,tchernin16} at $z=0.09$ \citep{owers11}. The cluster is located at the center of a supercluster \citep{einasto15,gramann15} and is characterized by ongoing accretion around its virial radius, as shown by the detection of several smaller structures \citep{owers11}. The intense dynamical activity is thought to be responsible for large-scale gas motions in the main cluster \citep{rossetti13}, leading to a highly elliptical X-ray morphology along the NW-SE axis. In this paper, we present the results of a deep \emph{Chandra} observation of the infalling group in the outskirts of Abell 2142 \citep[hereafter E14]{eckert14b}. This structure, located $\sim1.3$ Mpc from the core of A2142 perpendicular to the cluster major axis, exhibits a very long X-ray trail extending for at least 700 kpc. The group is in an advanced state of disruption, with less than 5\% of its total gas mass remaining within the core of the structure (the ``tip''). For these reasons, the infalling group of A2142 is an ideal target for studying the virialization of the accreting gas and placing constraints on the properties of the intracluster plasma.

The paper is organized as follows. In Sect. \ref{sec:analysis} we describe the data reduction procedure. In Sect. \ref{sec:tip} we present high-resolution imaging spectroscopy of the tip, which allows us to determine the motion and geometry of the core of the group. In Sect. \ref{sec:tail} we analyze the morphology and the spectral properties of the stripped gas, and in Sect. \ref{sec:ps} we study the power spectrum of surface-brightness fluctuations in the tail and set constraints on turbulence and transport processes following the method outlined in \citet{gaspari13,gaspari14}. We discuss our results in Sect. \ref{sec:disc}.

Throughout the paper, we assume a $\Lambda$CDM cosmology with $\Omega_m=0.3$, $\Omega_\Lambda=0.7$ and $H_0=70$ km/s/Mpc. At the redshift of A2142, this corresponds to $1^{\prime\prime}=1.7$ kpc. 

\section{Data analysis}
\label{sec:analysis}

The data were acquired in late 2014 and early 2015 during three different observations (Observation ID 17168, 17169, and 17492; PI: Eckert) for a total exposure time of 200 ks with ACIS-S. We analyzed the data using CIAO v4.8 and CALDB v4.7.2 and reprocessed the event files using the CIAO task \texttt{chandra\_repro}. We used the \texttt{fluximage} tool to extract ACIS-S images in the [0.5-2.0] keV band together with the corresponding exposure maps. The three datasets were then combined using the CIAO tool \texttt{dmmerge}. To estimate the background level, we used a collection of blank-sky pointings to extract a blank-sky image, which was then renormalized by the ratio of the [9.5-12] keV fluxes measured in the blank-sky data and in our observations, following the method outlined in \citet{HM06}. 

\begin{figure}
\hbox{\includegraphics[width=0.5\textwidth]{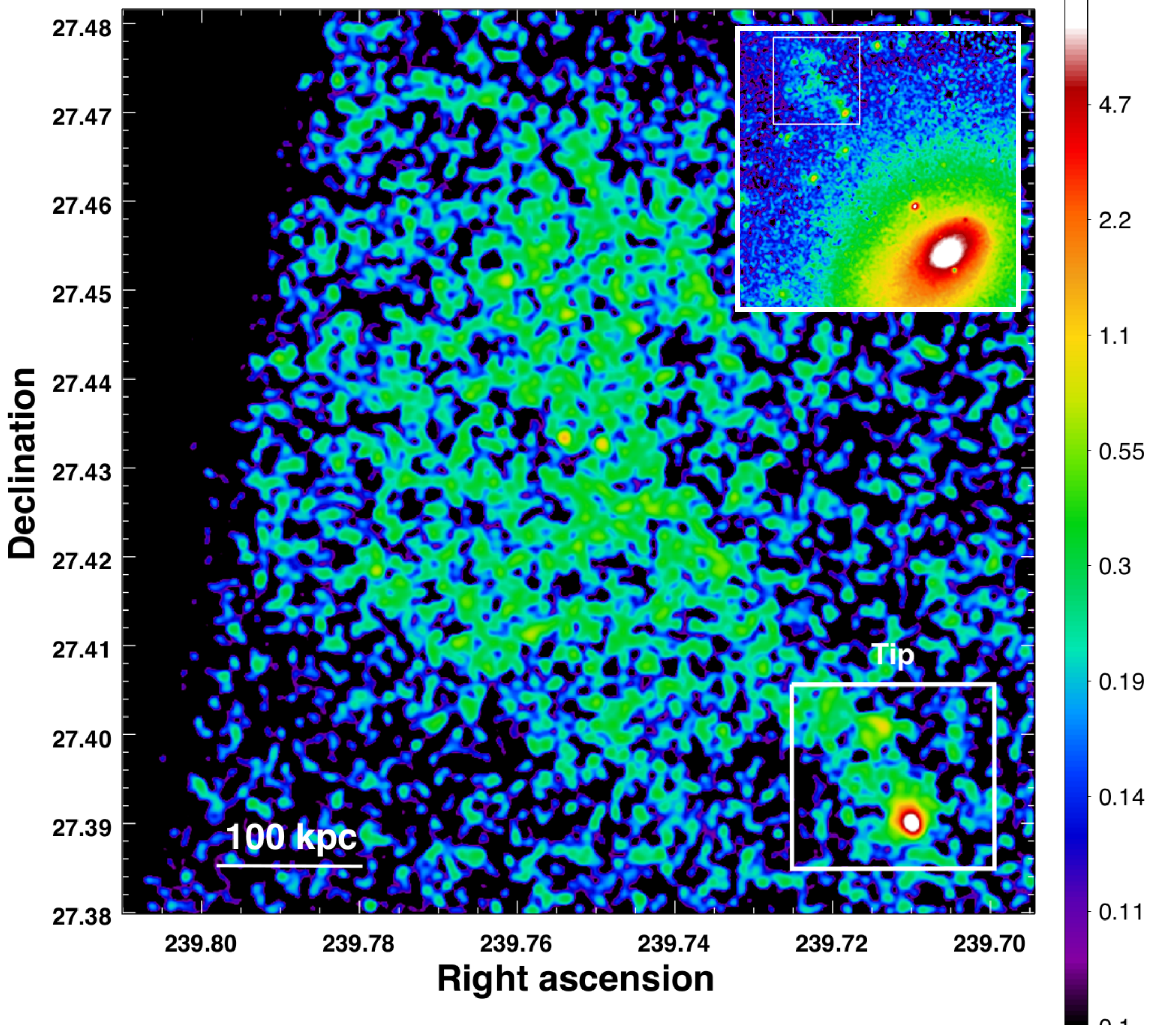}}
\caption{Raw \emph{Chandra}/ACIS-S image (in unit of counts) of the A2142 infalling group in the [0.5-2] keV band smoothed with a fixed Gaussian of 3 arcsec width. The white box in the bottom right corner shows the location of the ``tip'' (see Fig. \ref{fig:tip}). The inset in the top right corner shows the location of the accreting group compared to the core of the main cluster.}
\label{fig:raw}
\end{figure}

Point sources were extracted using the CIAO tool \texttt{wavdetect} down to a detection limit of ten counts per source, corresponding to a [0.5-2] keV flux limits of $4\times10^{-16}$ erg/cm$^2$/s. The corresponding areas were excised from the event files. The final image smoothed by a $3^{\prime\prime}$ Gaussian containing only the diffuse emission and combining all three observations is shown in Fig. \ref{fig:raw}. 

The diffuse emission from the structure is clearly detected in a region extending over several hundred kpc, and it can be subdivided into two main regimes: the tip around the position of the galaxies, inside which the gas is gravitationally bound to the group's halo, and the tail, which is composed of gas that has already been stripped from its original halo. In the following we describe the properties of these two regions separately.

\section{The tip}
\label{sec:tip}

\subsection{Morphology}

\begin{figure*}
\hbox{\includegraphics[width=0.49\textwidth]{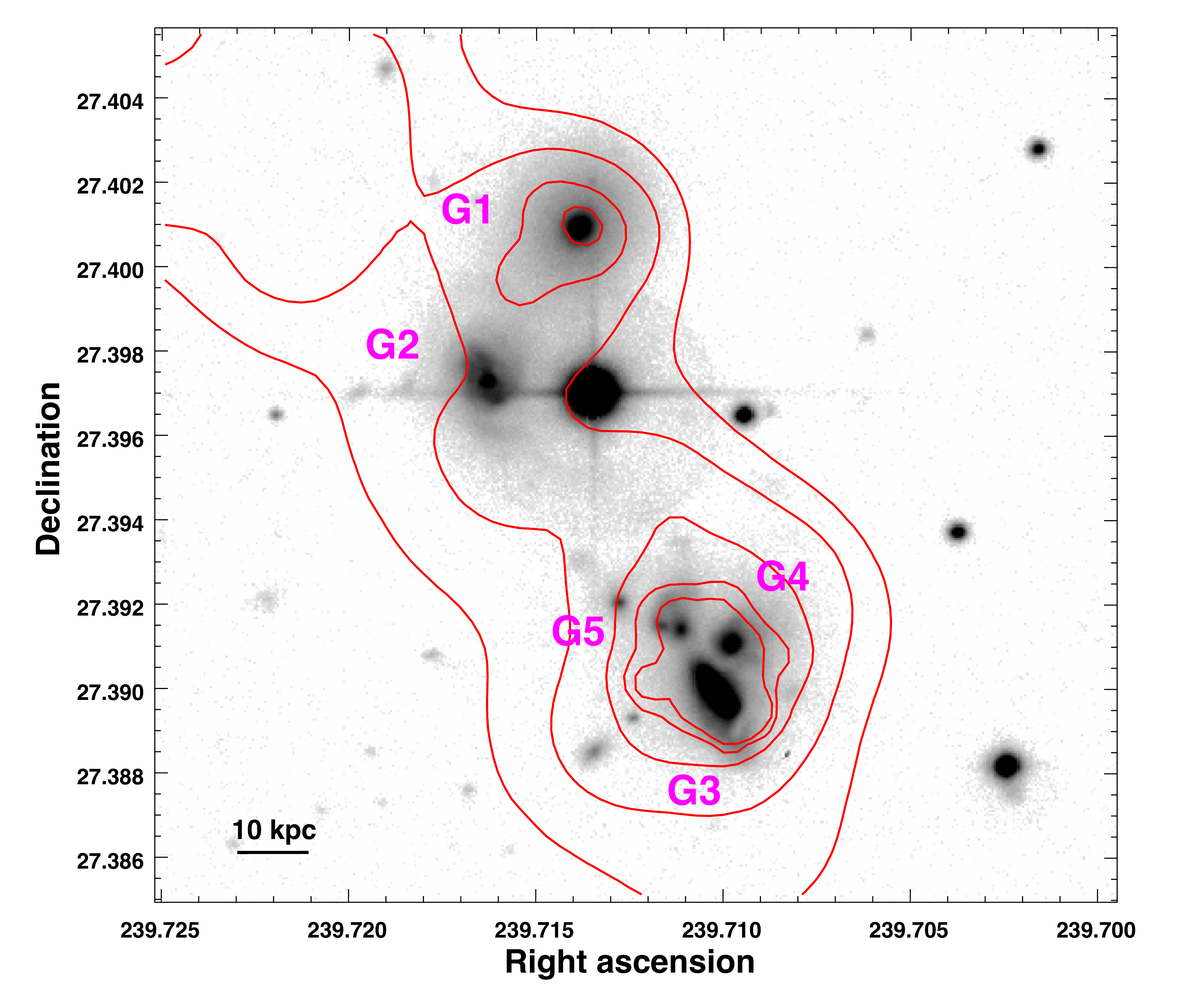}\includegraphics[width=0.52\textwidth]{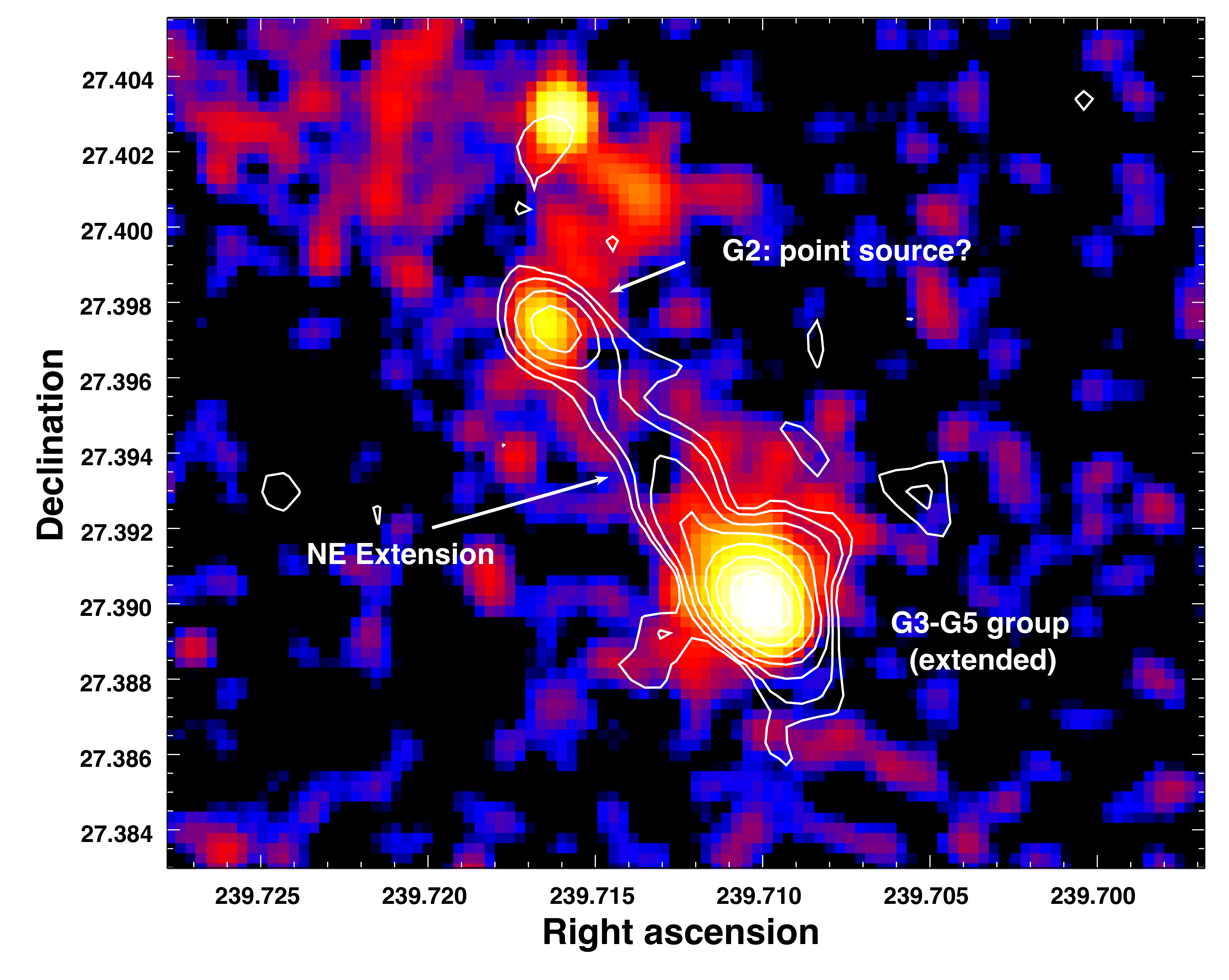}}
\caption{Tip of the structure. \emph{Left:} CFHT g-band image with X-ray contours in red. The galaxies discussed in E14 are highlighted in magenta. \emph{Right:} ACIS-S image in the [0.5-2] keV range including the point source, with GMRT radio contours at 610 MHz overlaid. Obvious radio features are highlighted in white.}
\label{fig:tip}
\end{figure*}

The tip of the structure is characterized by a very complex morphology (see Fig. \ref{fig:tip}). All the galaxies identified in Fig. 7 of E14 show pronounced X-ray emission. In Fig. \ref{fig:tip} we show the CFHT $g$-band image of the region with \emph{Chandra} contours overlaid. The five bright galaxies in this region are labeled G1 through G5, following the terminology of E14. The \emph{Chandra} image in the same region (including point sources) is shown for comparison, with GMRT radio contours at 610 MHz overlaid (Venturi et al. subm.). In the following we provide a description of the multiwavelength properties of these galaxies.

\begin{itemize}
\item G3-G5 is an interacting group of three morphologically disturbed galaxies with a peculiar velocity of $\sim1,000$ km/s with respect to the cluster mean \citep{owers11}.  The spectra of these galaxies are characterized by strong optical emission lines that indicate AGN and star formation activity (see E14). The brightest X-ray structure by far in the tip is the G3-G5 group, which exhibits extended emission on a scale of $\sim20$ arcsec, much larger than the \emph{Chandra} point-spread function
(PSF). The morphology of the diffuse emission of the G3-G5 group is elongated toward the NE, aligned with the direction of the tail. On the other hand, the surface brightness drops sharply toward the SW, corresponding to the direction of the cluster core. This suggests the presence of a possible edge. 

\item G2 is a face-on spiral galaxy with a peculiar velocity of -2,400 km/s. The X-ray source coincident with G2 is compact and consistent with a point source. The galaxy was found to host an AGN based on the [OIII]/H$\beta$ ratio (see E14). G2 is also a compact radio source, as shown in the right panel. Thus we identify this source as an AGN.

\item G1 is a passive elliptical galaxy that we identified in E14 as the probable dominant galaxy of the infalling group because of its elliptical morphology. This interpretation was supported by a low-significance substructure ($2\sigma$) in the velocity distribution of the galaxies in the region with a peculiar velocity consistent with the peculiar velocity of G1 (-231 km/s, see Fig. 8 of E14). The X-ray emission associated with this galaxy is diffuse, but weak. It shows no elongation in the direction of motion and is not perfectly aligned with the direction of the tail. No radio emission is associated. The bright source located just NE of G1 is point-like and has no obvious optical counterpart, thus we conclude that it is a background object. 

\item The radio image shows a blend of radio galaxies with a position consistent with G3 and G4 (see the right-hand panel of Fig. \ref{fig:tip}). We detect an extension of the radio emission toward the NE direction with a significance at least ten times the local r.m.s. This configuration is consistent with a head-tail morphology originating from one of the galaxies in the G3-G5 group, which again implies a motion in the SW direction toward the cluster core. High-resolution radio data at another frequency are necessary to measure the spectral index across the head-tail structure. The point-like radio source associated with G2 is probably a chance superposition with the group. No radio emission is observed coincident with the position of G1.
\end{itemize}

\subsection{Spectral properties}

To investigate the spectral properties of the gas within the tip, we extracted the X-ray spectrum in several regions using the CIAO tool \texttt{specextract} and studied their spectral properties. Spectral fitting was performed using {\sc Xspec} v12.9. The local background was calculated in a neighboring region and subtracted from the data. The source was modeled with the thin-plasma emission code APEC \citep{smith01xray} absorbed by the Galactic $N_H$, which was fixed to the 21cm value of $3.8\times10^{20}$ cm$^{-2}$ \citep{kalberla05}. The C-statistic was used to estimate the best-fit parameters \citep{cash79}.

\begin{figure*}
\hbox{\includegraphics[width=0.5\textwidth]{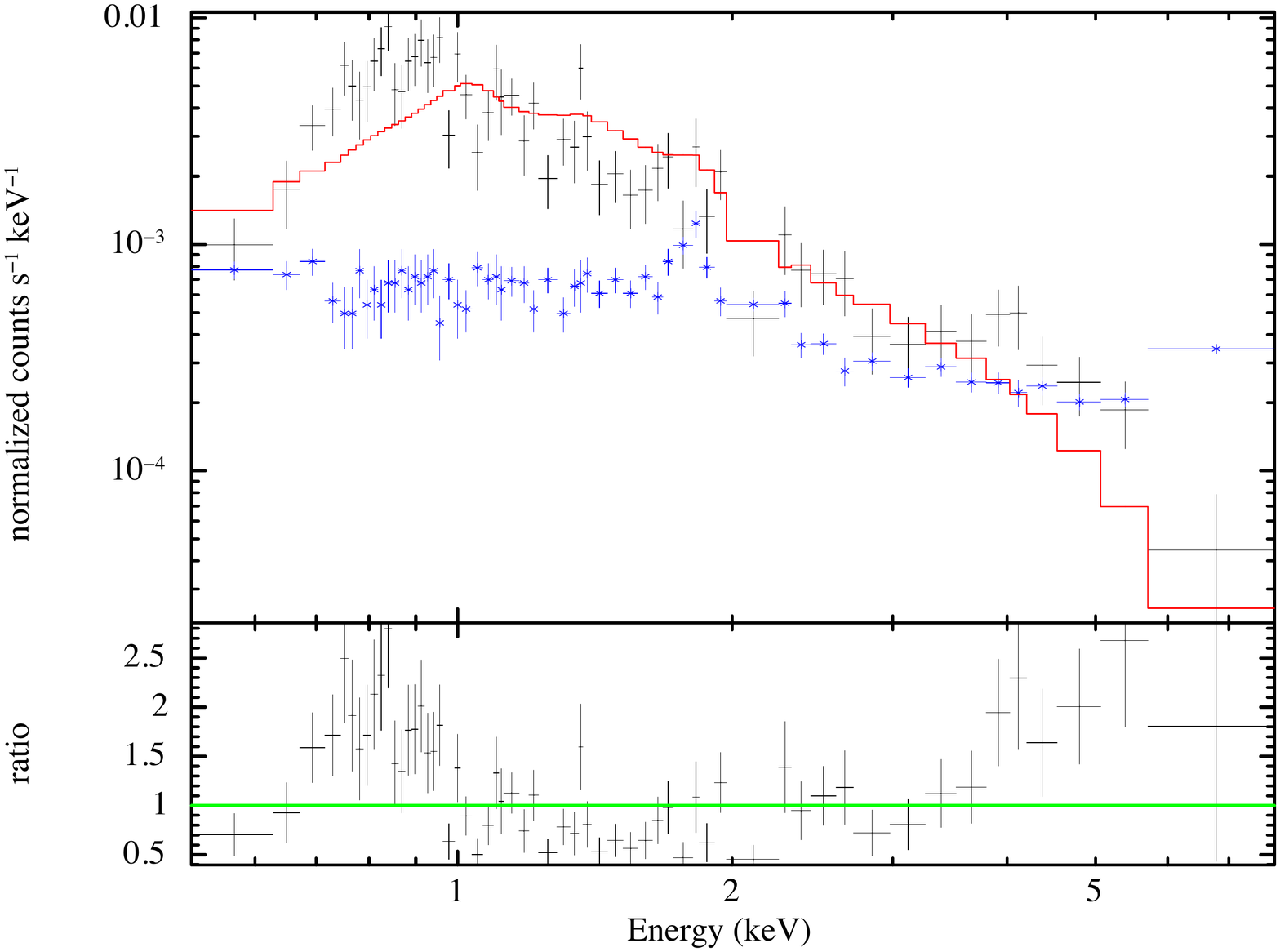}\includegraphics[width=0.5\textwidth]{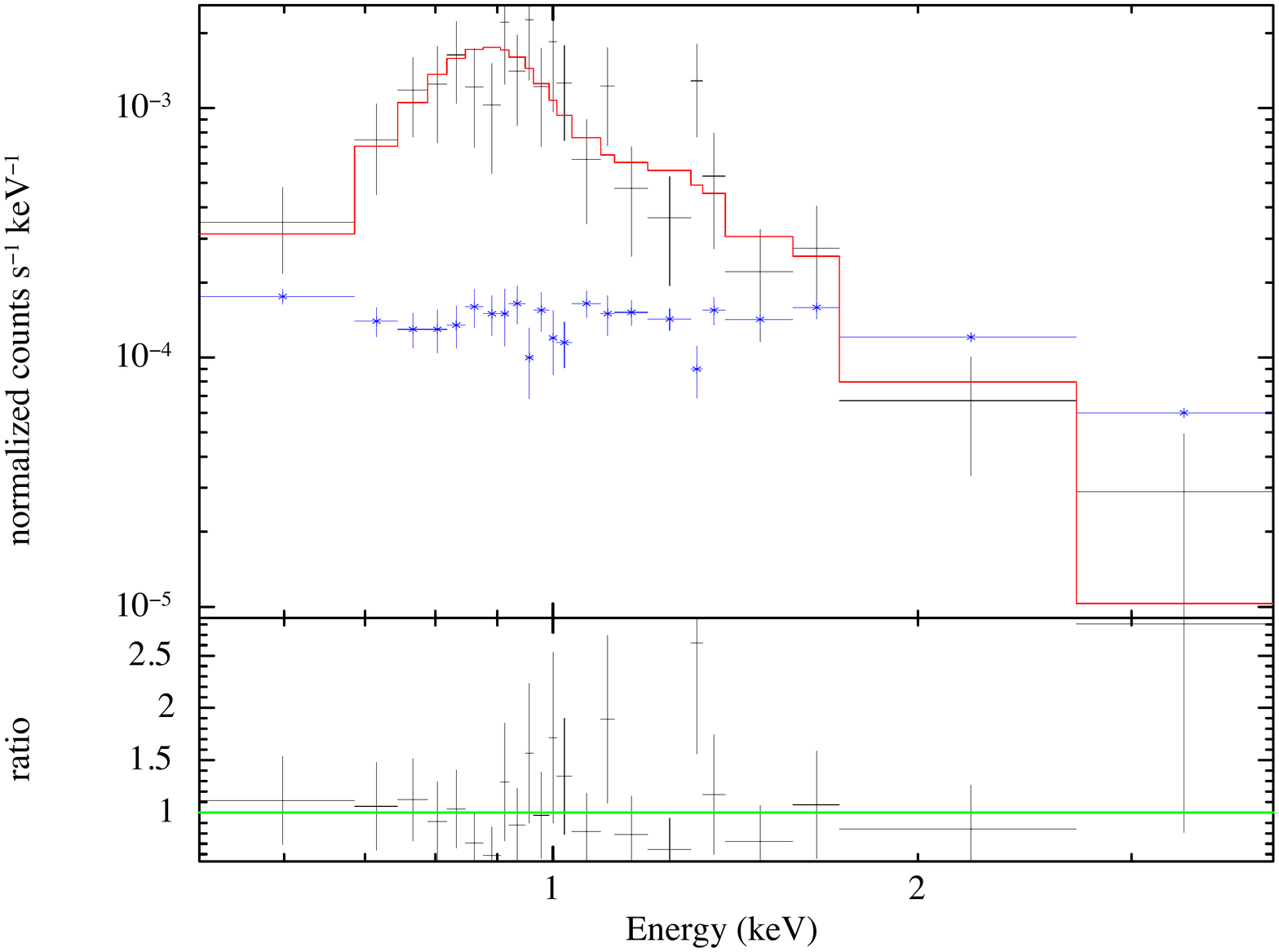}}
\caption{ACIS-S spectrum of the G3-G5 system, fit with a single-temperature APEC model. On the left, the spectrum of the full G3-G5 group is shown (1 arcmin aperture), while on the right an annulus excluding the G3 and G4 galaxies has been used. The bottom panels show the residuals from the fit. In both cases, the background level is shown in blue.}
\label{fig:g3g5spec}
\end{figure*}

In the left-hand panel of Fig. \ref{fig:g3g5spec} we show the spectrum of the source fit with a single-temperature APEC model with metal abundance fixed to 0.3$Z_\odot$ and absorbed by the Galactic $N_H$. As shown in the fit residuals, this model provides a poor fit to the data (C-stat=$225.8/115$ d.o.f.) and a high temperature of 3.2 keV. A better fit is achieved when adding a hard power law to the model with a photon index fixed to the value $\Gamma=1.6$ (C-stat=$130.6/114$ d.o.f.). We then obtain a more realistic temperature of $0.87\pm0.08$ keV. Similarly, a single power law cannot reproduce the data. Thus, the spectrum is likely contaminated by one or several AGN that may be responsible for the associated radio emission. 

To investigate this possibility, we extracted an image of the tip in the [2-7] keV band, where the emission is expected to be dominated by the AGN. This allowed us to confirm that the hard-band emission is dominated by two point-like sources, whose positions are consistent with the nuclei of G3 and G4. The luminosities of these AGN in the [2-10] keV band are $(2.4\pm0.3)\times10^{42}$ erg/s and $(3.0\pm0.4)\times10^{42}$ erg/s for G3 and G4, respectively. We then extracted the X-ray spectrum of an annular region excluding these two point sources and again fit the resulting spectrum (see the right-hand panel of Fig. \ref{fig:g3g5spec}). The spectrum of the annulus is well described by a single-temperature APEC model with $kT=0.98_{-0.08}^{+0.06}$ keV and a metal abundance $Z/Z_\odot=0.18_{-0.08}^{+0.13}$. This confirms that the hot gas associated with the G3-G5 group has a temperature on the order of 1 keV, which is similar to the temperature of $\sim1.3$ keV measured in the tail (see E14). The gas associated with G3-G5 is thus consistent with being the core of the accreting group.\\

We also extracted the spectrum of the weak diffuse emission associated with the G1 elliptical. The spectrum of this region is soft, and it can be well represented by an absorbed APEC model with $kT=0.92_{-0.25}^{+0.40}$ keV. The integrated luminosity of this source is very low ($L_{[0.5-2]}\sim5\times10^{40}$ ergs/s).

\subsection{Leading edge}

As can be seen in the right-hand panel of Fig. \ref{fig:tip}, the surface brightness of the G3-G5 group sharply drops in the SW direction, which suggests the presence of a leading edge. To investigate this possibility, we extracted a surface-brightness profile from the peak of the G3-G5 group in the sector with position angles 210$^\circ$-30$^\circ$ using {\sc Proffit} v1.3 \citep{ccbias1}. The surface-brightness profile is very sharp and drops by nearly two orders of magnitude 8 arcsec (14 kpc) SW of the surface-brightness peak. 

\begin{figure}
\includegraphics[width=0.5\textwidth]{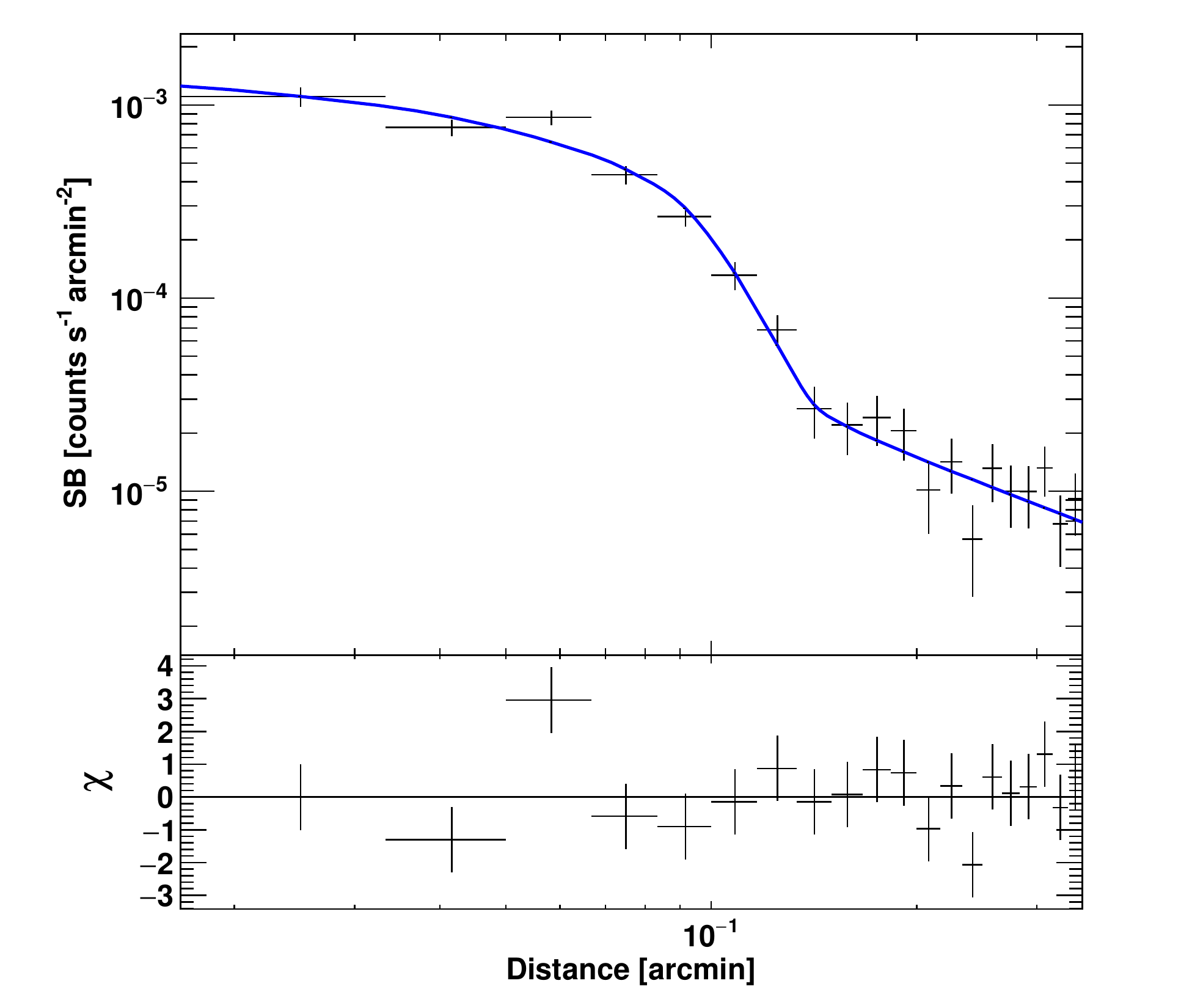}
\caption{Surface-brightness profile from the peak of the G3-G5 group across the leading edge (PA 210-30$^\circ$). The blue curve shows the best fit to the data using a broken power-law model projected along the line of sight. The residuals are shown in the bottom panel.}
\label{fig:edge}
\end{figure}

To investigate whether the sharp drop could be due to a density discontinuity, we fit the output profile with a broken power-law projected along the line of sight \citep{owers09,eckert16}. The profile and best-fit model are shown in Fig. \ref{fig:edge}. A clear surface-brightness edge is observed 8 arcsec SW of the peak, coincident with the possible leading edge identified in Fig. \ref{fig:tip}. Given that the temperature inside the group is much lower than the temperature of the surrounding plasma ($\sim6$ keV), this feature is a cold front. The broken power-law provides a fairly good representation of the data ($\chi^2=21.5/18$ d.o.f.) and it returns a very high density jump $n_{\rm in}/n_{\rm out}=4.5_{-1.0}^{+1.6}$ at the edge. We note, however, that this value is rendered unreliable by the AGN in G3 (see above), which provides a substantial contribution to the flux measured in Fig. \ref{fig:tip}. Nonetheless, our analysis shows a cold front at the interface between the moving group and the surrounding ICM and proves that the G3-G5 group is moving almost radially in the direction of the center of A2142.

\subsection{Interpretation}

To summarize, the leading edge, the associated thermal X-ray emission, the extension of the emission in the direction of the tail (see Fig. \ref{fig:tip}), and the head-tail morphology of the radio emission are strong arguments in favor of the association of the G3-G5 structure with the core of the infalling group. Conversely, the very weak diffuse emission at the position of G1 and the lack of apparent connection between G1 and the X-ray tail indicate that G1 is probably unrelated to the diffuse X-ray structure. Considering that groups of a mass of $\sim3\times10^{13}M_\odot$ typically have an X-ray luminosity on the order of $10^{43}$ ergs/s \citep[e.g.,][]{lovisari15,giles16}, the X-ray luminosity of the G1 galaxy ($\sim5\times10^{40}$ ergs/s) is far too low to be attributed to the core of the group. Such a luminosity is typical of X-ray coronae of elliptical galaxies in clusters \citep{sun07b,vija15}, which suggests that this galaxy has accreted onto the cluster a long time ago. 

For all the above reasons, we conclude that the G3-G5 interacting group is associated with the core of the X-ray structure. The wealth of active phenomena observed in these galaxies (AGN activity, head-tail radio galaxy, star formation activity) are probably induced by the interaction with the main cluster. Given its large difference in line-of-sight velocity with the other members, the G1 galaxy is likely seen in projection and is unrelated to the X-ray structure, at odds with the interpretation put forward in E14. 

\section{The tail}
\label{sec:tail}

\begin{figure}
\centerline{\includegraphics[width=0.5\textwidth]{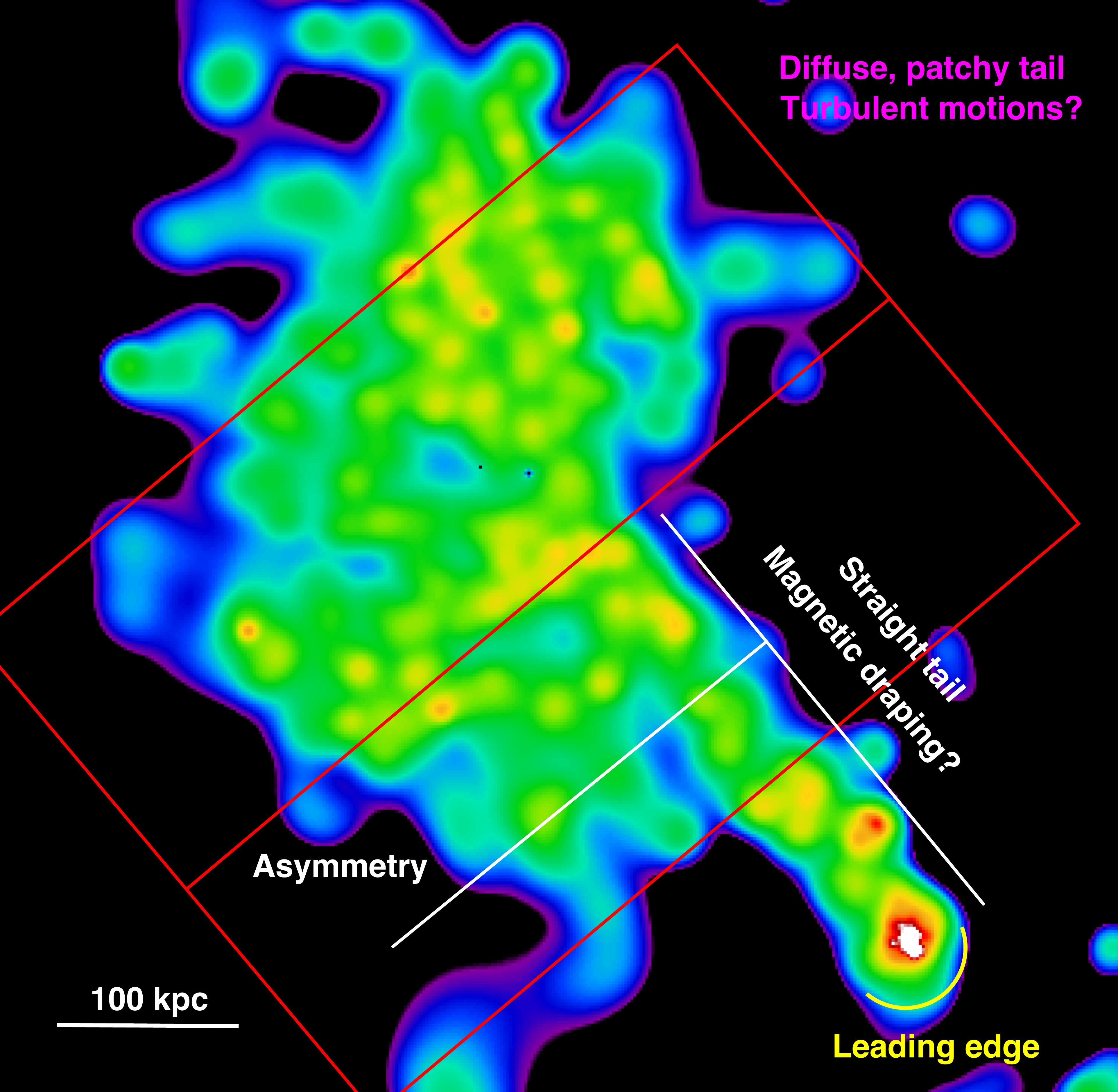}}
\caption{Adaptively smoothed ACIS-S image of the tail in the [0.5-2] keV band. A number of clear morphological features are identified in the image. The red boxes indicate the regions we
used to extract the surface-brightness profiles shown in Fig. \ref{fig:boxes}.}
\label{fig:features}
\end{figure}

The deep \emph{Chandra} observation of the tail reveals an amazing wealth of structures induced by the motion of the structure. In Fig. \ref{fig:features} we show the adaptively smoothed background-subtracted image of the tail with a number of features identified. The blank-sky background was subtracted from the original image (see Sect. \ref{sec:analysis}) and the resulting image was adaptively smoothed using \texttt{asmooth} \citep{ebeling06} with a target signal-to-noise
ratio of 5.0 to enhance the diffuse emission.

\subsection{Morphology}

The tail can be subdivided into two characteristic parts: a straight, asymmetric regime closest to the tip, and an irregular, morphologically complex region located beneath. Overall, the width of the structure increases with distance from the tip, as one would expect for the stripping of progressively larger gas shells, forming a three-dimensional geometry resembling a cone oriented close to the plane of the sky. 

\begin{figure}
\centerline{\includegraphics[width=0.5\textwidth]{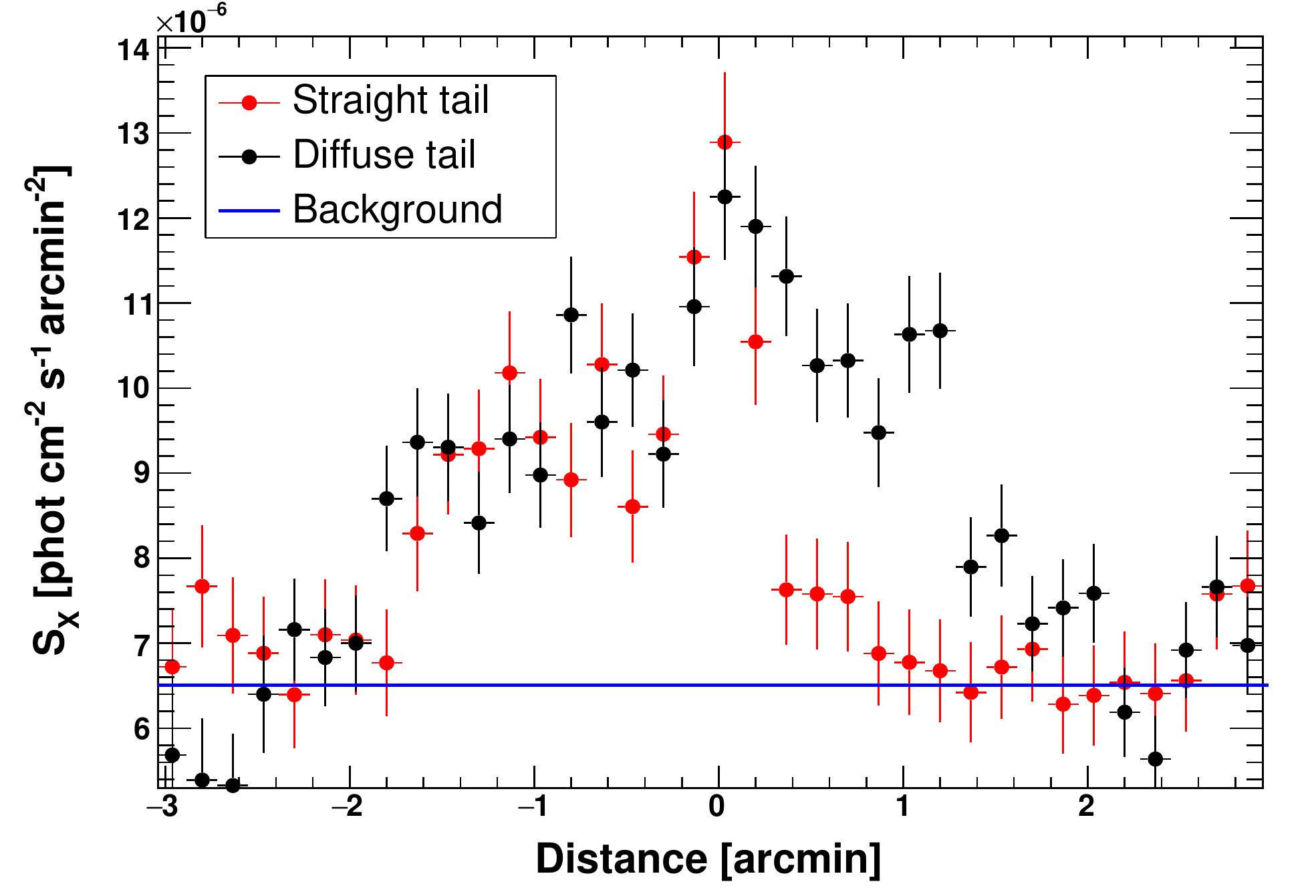}}
\caption{ACIS-S surface-brightness profiles cross the tail, along the two boxes defined in Fig. \ref{fig:features}. The zero-point on the X axis is set along the axis of motion, while negative and positive values indicate the surface brightness in the southeast and northwest direction, respectively.}
\label{fig:boxes}
\end{figure}

The region closest to the tip, which is made of gas that was stripped at late times, appears straight over a scale of $\sim250$ kpc. Interestingly, beyond the innermost $\sim100$ kpc the structure of the tail appears strongly asymmetric perpendicular to the group motion. While NW of the tail the emission appears to drop sharply, in the SE direction a patch of diffuse emission can
be observed that
was already present in the \emph{XMM-Newton} data (see Fig. 5 of E14). To confirm this statement, we extracted surface-brightness profiles across two boxes perpendicular to the main axis of the structure at a mean distance of 200 kpc and 370 kpc from the tip. The corresponding brightness profiles are shown in Fig. \ref{fig:boxes}, where zero distance corresponds to the position of the straight tail. We can see that the brightness profile perpendicular to the straight tail differs significantly between the two cases. While at larger distance from the core of the group the emission is more symmetric with respect to the main axis, in the region located 200 kpc from the tip we see a strongly asymmetric behavior and a sharp edge in the northwest direction. This asymmetry might be explained by projection effects if the group is moving at a substantial velocity along the line of sight and with a non-zero impact parameter, similar to cometary tails. This is consistent with the peculiar velocity of the G3-G5 structure (700-1,200 km/s), which indicates that there is a significant line-of-sight component to the motion. 

 
Beyond 300 kpc from the tip, the tail opens up and reaches a width of more than 300 kpc. This observation is consistent with the idea that was put forward in E14, which is that the gas located in the outermost regions of the tail was originally located in the outer regions of the group and was stripped first because of its lower thermal pressure. The surface-brightness distribution of this region is patchy and irregular, which is reminiscent of chaotic motions caused by turbulence and/or mixing. The morphology of the tail in this region cannot be described any more by a conical geometry, which suggests that the gas is no longer confined and is expanding within the surrounding ICM.

\begin{figure*}
\hbox{\includegraphics[width=0.5\textwidth]{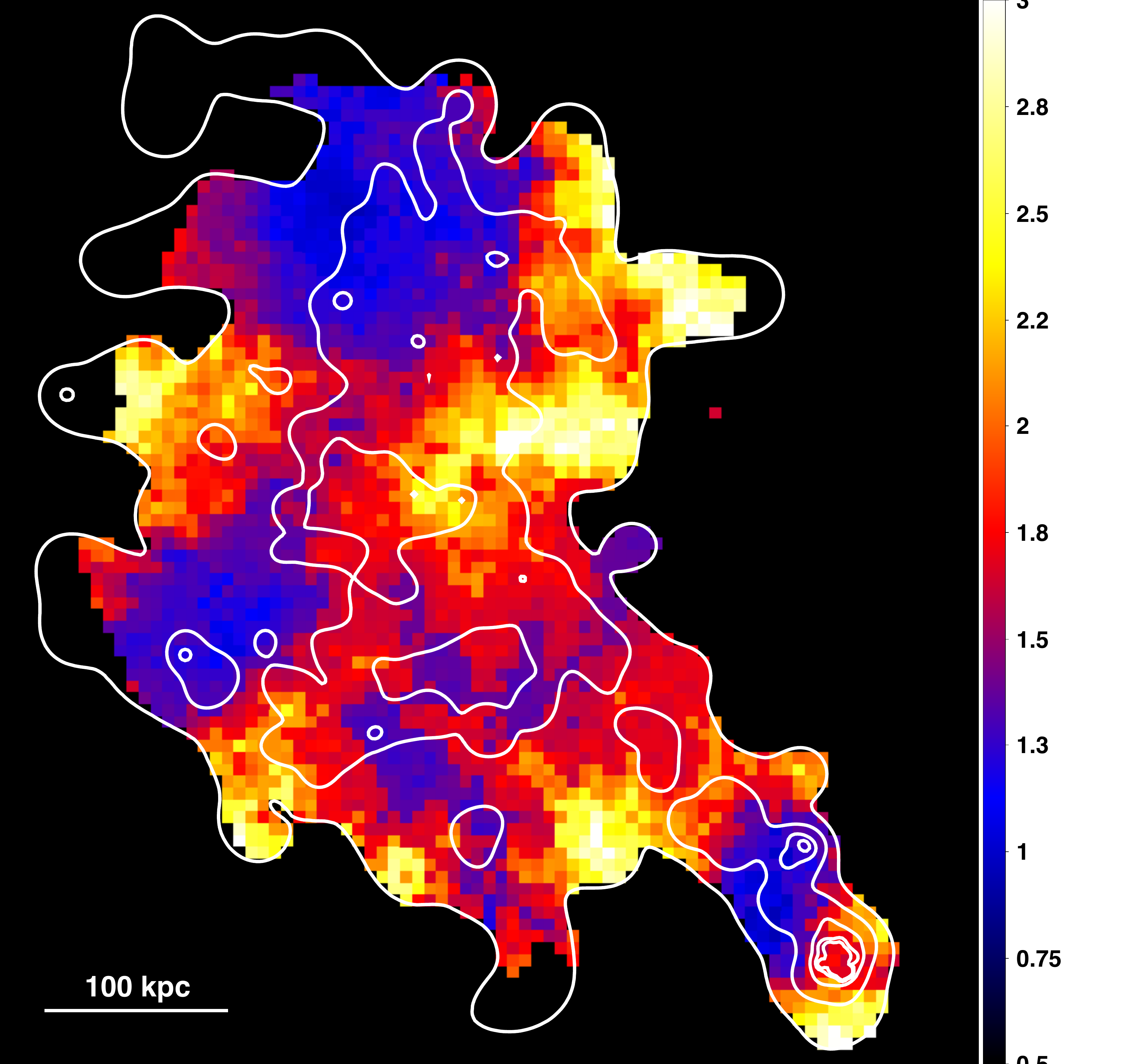}\includegraphics[width=0.5\textwidth]{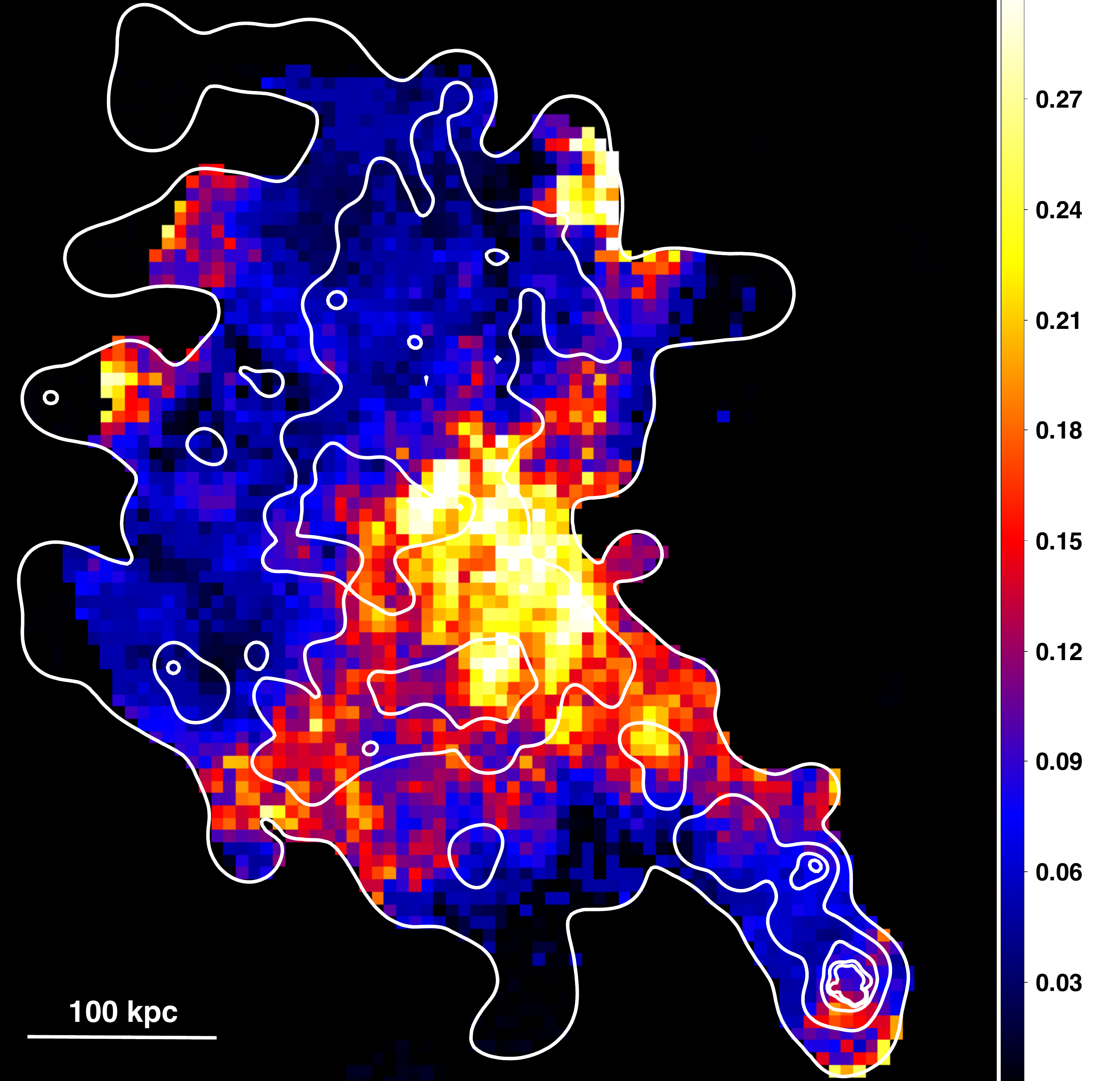}}
\caption{ACIS-S temperature map (left) and metal abundance map (right) obtained by spectral fitting in circular regions containing 500 source counts. The cyan contours show X-ray isophotes in the [0.5-2.0] keV band.}
\label{fig:spectroimaging}
\end{figure*}

\subsection{Spectro-imaging analysis}

We extracted temperature and metal abundance maps in the tail following the method of \citet{owers13,owers14}. Briefly, we produce a background-subtracted combined image in the [0.3-3] keV range from which point sources are removed and that is binned by a factor of 8 (i.e., with a pixel size of $3.94^{\prime\prime}\times3.94^{\prime\prime}$). This image is used to define a circular region centered at each pixel with the radius set such that the region contains a set number of source counts, $N_{src}$. To determine the $N_{src} $ required to obtain temperature measurements with $10\%$ relative uncertainty ($90\%$ uncertainties), we simulated an absorbed MEKAL model with $N_H=3.8\times10^{20}\,{\rm cm}^{-2}$, $kT=1.5$\,keV and abundance $Z=0.25 Z_{\odot}$ at the cluster redshift. We find that $N_{src}=500$ is sufficient for this purpose. We extracted source and background spectra from the circular regions centered at each pixel in the binned image. Response files were extracted on a coarser $7.87^{\prime\prime}\times7.87^{\prime\prime}$ grid. The extracted spectra were fitted in {\sc Xspec} with an absorbed MEKAL model, where the temperature, normalization and abundance are free to vary. The redshift and hydrogen column density were fixed to $z=0.09$ and $N_H = 3.8\times10^{-20}\,{\rm cm}^{-2}$, respectively.  When we proceed in this way, the recovered values of temperature and metallicity are obviously correlated; the correlation length ranges from $\sim 50$\,kpc to $\sim 90$\,kpc.

The output temperature and abundance maps are shown in Fig. \ref{fig:spectroimaging} with surface-brightness contours overlayed. The temperature of the tail appears almost uniform given the uncertainties and varies in the range 1-2 keV, in agreement with the \emph{XMM-Newton} values presented in E14. We recall that the temperature of the surrounding plasma is expected to be $\sim5$ keV \citep{tchernin16}. Overlying brightness contours onto the temperature and metallicity maps show that the gas near the tip is cool ($\sim1$ keV), while there are patches of similarly cool gas in the tail interspersed with $\sim2$ keV regions. A region of $\sim1.5$ keV gas with a metallicity of $\sim0.15-0.2$ is located $\sim300$ kpc from the tip. Interestingly, this region is immediately followed by a drop in the measured metallicity, which agrees with the drop between sectors 2 and 3 reported in Table 2 of E14. To confirm this statement, we manually extracted the spectra of adjacent regions showing high and low metallicity in the map. In the former case, a clear bump around 1 keV associated with the Fe L line complex is observed, confirming the significant metallicity of the gas in this region. Conversely, the spectrum of the low-metallicity region appears featureless, and the best-fit metallicity is consistent with 0. 

Interestingly, we can see in Fig. \ref{fig:spectroimaging} that the border between the low- and high-metallicity regions occurs close to the transition between the straight tail and the irregular diffuse tail. The findings presented here are discussed in more detail in Sect. \ref{sec:disc}.


\section{Power spectrum of the density fluctuations}
\label{sec:ps}

To study the microphysics of the gas in the tail further, we calculated the power spectrum of surface-brightness fluctuations in the tail. Recent works \citep{gaspari13,gaspari14,zhu14,hofmann16,khatri16} have shown that the amplitude of surface-brightness fluctuations is proportional to the Mach number of turbulent motions in the plasma, while the slope of the power spectrum is sensitive to dissipative processes, in particular thermal conduction \citep{gaspari13}. The flat surface-brightness distribution in the tail, especially in the region beyond 200 kpc where the straight tail disappears, makes this object an excellent target for studying the surface-brightness fluctuations and link them with the microphysics. We verified that the complex 3D geometry of the structure does not affect the slope of the measured power spectrum by simulating projected 2D power spectra in a conical geometry with different projection angles, demonstrating that for inclination angles of 30$^\circ$ or smaller the slope is left unaltered on small scales (see Appendix \ref{app:conicalps}).

\subsection{Method}

We used the modified $\Delta$-variance method proposed by \citet{arevalo12} to extract the power spectrum of surface-brightness fluctuations. This method is well suited to extract the power spectra of data with gaps and non-periodic boxes. In more detail, to calculate the power at scale $\sigma$, the raw image $I$ is convolved with a Mexican-hat filter (implemented as the difference of two Gaussians) and corrected for border effects by dividing the convolved images by the convolved exposure maps $E$, 

\begin{equation}I_\sigma = \left(\frac{G_{\sigma_1}\circ I}{G_{\sigma_1}\circ E}-\frac{G_{\sigma_2}\circ I}{G_{\sigma_2}\circ E}\right)\times E,\label{eq:mh}\end{equation}

\noindent where $G_\sigma$ is the Gaussian filter at scale $\sigma$ and $\sigma_{1}, \sigma_{2}$ differ by a small amount, that is,
by $\sigma_{1}=\sigma/\sqrt{1+\epsilon}$, $\sigma_{2}=\sigma\sqrt{1+\epsilon}$ with $\epsilon\ll1$. Following \citet{arevalo12}, we set $\epsilon=10^{-3}$. The convolved image $I_{\sigma}$ selects fluctuations with wave number $k=1/{\sqrt{2\pi^2}\sigma}$ and suppresses fluctuations on scales $\gg\sigma$ and $\ll\sigma$; thus the variance $V_{k}=\sum I_{\sigma}^2$ of $I_{\sigma}$ is proportional to the power $P(k)$ \citep[Eq. 17 of][]{churazov12}.

We convolved the ACIS-S image with the modified Mexican-hat filter (Eq. \ref{eq:mh}) for scales in the range $2-35$ pixels (where 1 pixel = 0.984 arcsec = 1.67 kpc). The G3-G5 group was masked during the convolution process to avoid leaking power to the region of the tail. Brightness fluctuations over the local background can be clearly observed in the convolved images, confirming that surface-brightness fluctuations are significant. To calculate the amplitude of the fluctuations, we defined a region that closely
traces the diffuse tail shown in Fig. \ref{fig:features}. We assumed that the underlying average surface-brightness profile is flat (as indicated in Fig. 4 of E14) and calculated the mean surface brightness $\mu_{S}$ in the region. In this case, the power spectrum of brightness fluctuations can simply be estimated by dividing the measured power by the squared mean,

\begin{equation}P_{2D}(k)=\frac{1}{\epsilon^2\pi k^2\mu_S^2}\frac{V_{k}}{\sum E^2}.\end{equation}

The characteristic 2D amplitude is then given by $A_{2D}=\sqrt{P_{2D}(k)2\pi k^2}$. 

\begin{figure*}
\hbox{\includegraphics[width=0.5\textwidth]{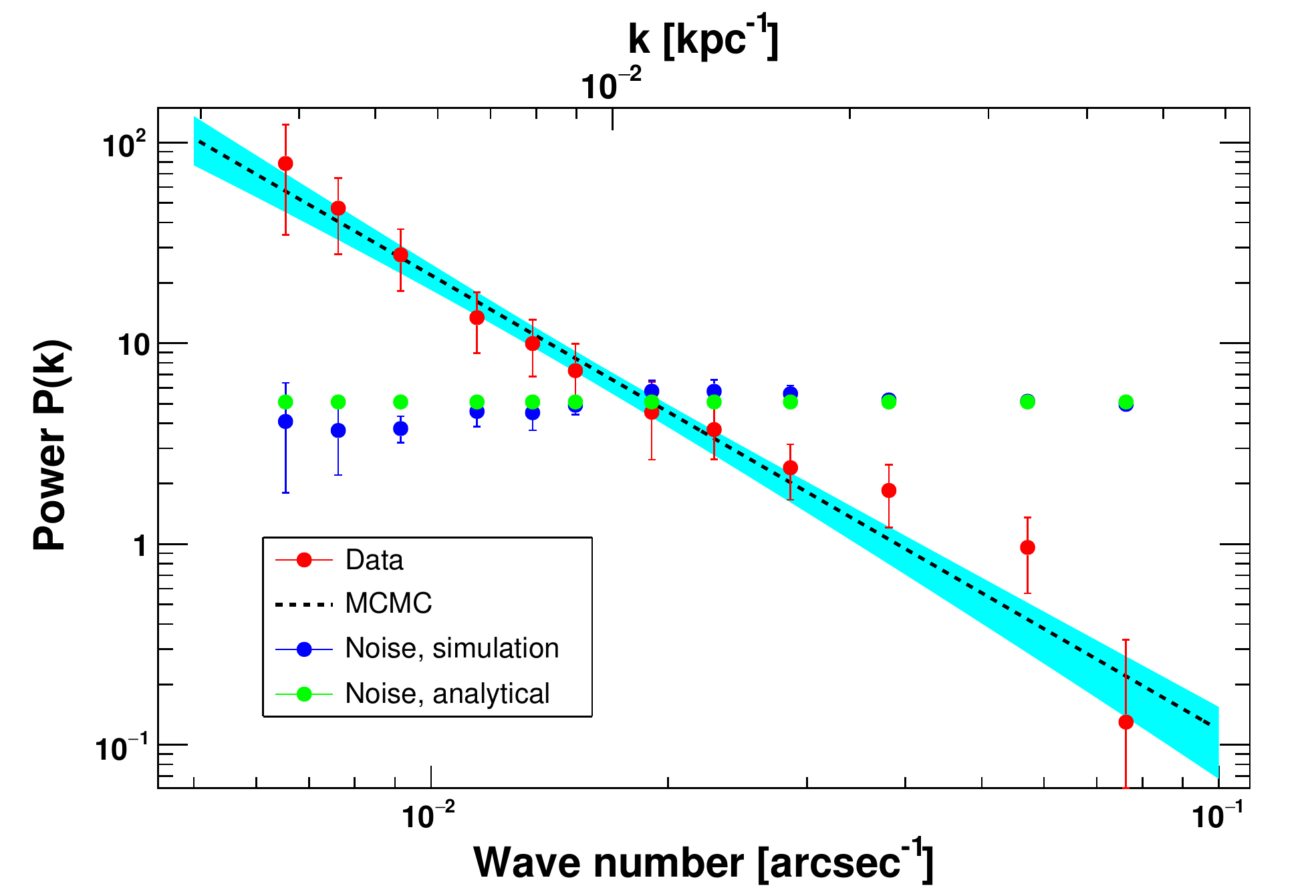}\includegraphics[width=0.5\textwidth]{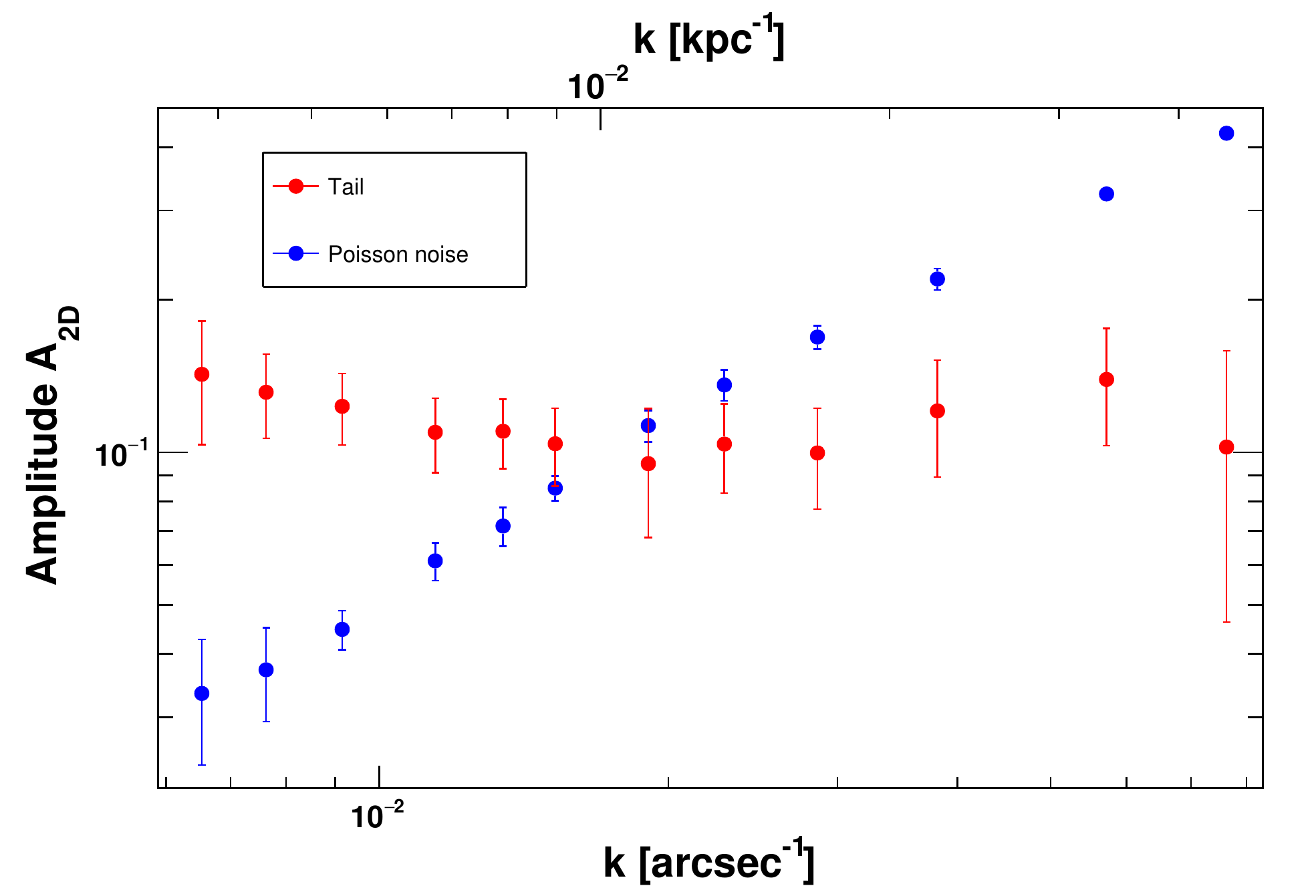}}
\caption{Power spectrum of surface-brightness fluctuations (red points) fit with a power law taking covariance into account. The dashed black line and cyan envelope denote the $68\%$ confidence region obtained using MCMC. The green and blue data points show the analytical expectation for the Poisson noise and the power spectrum recovered from a simulation including Poisson noise and unresolved point sources. \emph{Right:} Characteristic amplitude of the fluctuations (red) compared to the simulated Poisson noise (blue).}
\label{fig:ps}
\end{figure*}

\subsection{Estimating the photon shot noise}

Since the variance of Poisson random variables is equal to the measured value, the expected photon shot noise can be estimated analytically to be given by the formula \citep[see Eq. 18 of ][]{churazov12}

\begin{equation}P_{\rm Poisson}(k)=\frac{\sum I_{\rm raw}}{\mu_S^2\sum E^2}.\label{eq:poisson}\end{equation}

To test the validity of our method and our estimate of the Poisson shot noise, we simulated a Poisson image with a flat surface brightness equal to $\mu_S$, including vignetting effects and unexposed areas. In addition, unresolved point sources with a total flux below our detection limit can induce an additional flat term to the power spectrum. To take this effect into account, we used the $\log N-\log S$ of cosmic X-ray background sources from \citet{moretti03} and randomly added point-like sources with a flux in the range $10^{-17}-4\times10^{-16}$ ergs/cm$^2$/s (i.e., below our detection threshold, see Sect. \ref{sec:analysis}). The final simulated image thus includes both photon shot noise and the contribution from unresolved active galactic nuclei. We then calculated the power spectrum of the simulated image in the same way as for the data. The resulting power spectrum is flat and is consistent with the analytical expectation (Eq. \ref{eq:poisson}).

\subsection{Uncertainties and covariance matrix}

Given that the Fourier transform of the modified Mexican-hat filter (Eq. \ref{eq:mh}) is not a delta function \citep{arevalo12}, the power calculated on a given scale is correlated with that measured on different scales. This introduces a covariance between our data points that needs to be taken into account to determine the slope of the power spectrum. To estimate the uncertainties and calculate the covariance matrix, we split the region into 20 independent subregions and calculated the variance in each of the subregions. We then generated $N_r=10,000$ bootstrap realizations of the power spectrum by shuffling the 20 subregions with repetition, taking the mean of each realization, and calculating the variance of the distribution at each scale. The covariance matrix was then calculated from the 10,000 realizations of the power spectrum,

\begin{equation}\Sigma^2_{i,j}=\frac{1}{N_r}\sum_{\ell=1}^{N_r} (P_\ell(k_i)-P(k_i))(P_{\ell}(k_j)-P(k_j)),\end{equation} 

\noindent where $P_\ell(k_i),P(k_i)$ represent the $\ell^{\rm th}$ realization of the power spectrum and the mean value at scale $k_i$, respectively.

In Fig. \ref{fig:ps} we show the output power spectrum and characteristic amplitude of the fluctuations, subtracted for the photon shot noise. The uncertainties shown here are the square root of the diagonal elements of the covariance matrix. 

\subsection{Slope of the power spectrum}

As is shown in Fig. \ref{fig:ps}, the observed shape of the power spectrum resembles a power law. Thus, we modeled the power spectrum with a simple power law, 

\begin{equation}F(k)=P_{0}\left(\frac{k}{(\mbox{25 kpc})^{-1}}\right)^{-\alpha}.\label{eq:pl}\end{equation}

\begin{figure}
\hbox{\includegraphics[width=0.5\textwidth]{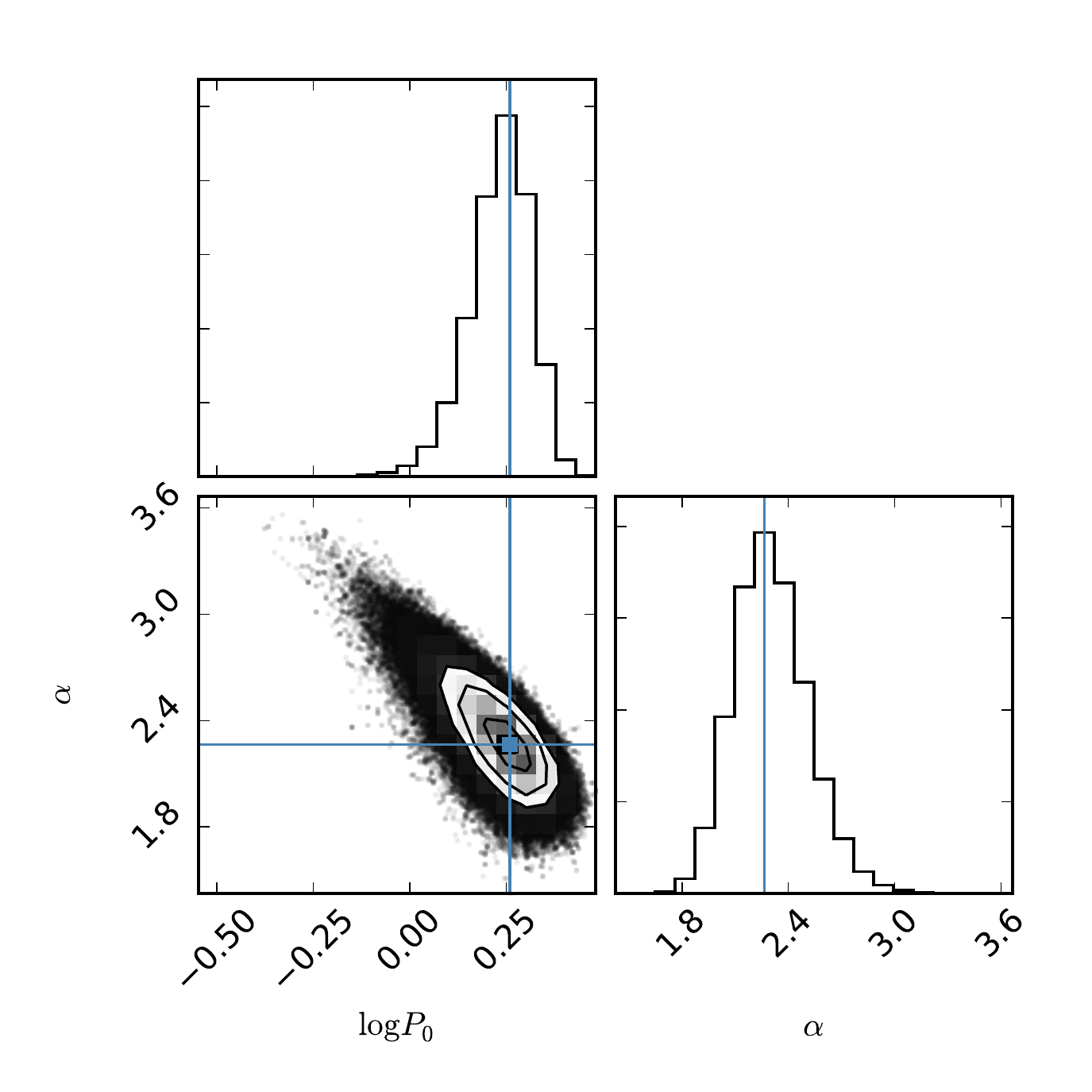}}
\caption{Posterior distribution of the parameters (see Eq. \ref{eq:pl}) obtained from 10,000 MCMC steps. The blue lines indicate the best-fit parameters $\alpha=2.28$, $P_{0}=1.74$.}
\label{fig:mcmc}
\end{figure}

\noindent We constructed a likelihood function taking the covariance between the data points into account,

\begin{equation} \log\mathcal{L}=-\frac{1}{2}\sum_{i,j=1}^N (P_{i}-F(k_i))(P_{j}-F(k_j))(\Sigma^2)^{-1}_{i,j}, \end{equation}

\noindent where $(\Sigma^2)^{-1}$ is the inverse of the covariance matrix. We then sampled the parameter space using the affine-invariant Markov chain Monte Carlo (MCMC) code \texttt{emcee} \citep{foreman-mackey13}. After a burn-in phase of 1,000 steps to ensure that the chains have converged, we performed 10,000 MCMC steps. The posterior distributions of the model parameters are shown in Fig. \ref{fig:mcmc}. 

Our final estimates for the parameter values are $\alpha=2.28_{-0.19}^{+0.22}$ and $P_0=1.74\pm0.30$. Therefore, the slope of the power spectrum is flatter than the Kolmogorov slope and it is only slightly steeper than 2.0, which implies a nearly constant characteristic amplitude for the fluctuations, as shown in the right-hand panel of Fig. \ref{fig:ps}. The observed amplitude is in the range $\delta S_X/S_X=0.1-0.14$ throughout the entire range of wave numbers accessible to this study. As shown in Appendix \ref{app:conicalps}, the conversion between $P_{2D}$ and $P_{3D}$ is rendered complicated by the complex morphology of the source. For a conical geometry and inclination angles in the range $0-30^\circ$, our simulations show that the normalization of the projected power spectrum changes by a factor of 4, that is, the amplitude of the 3D density fluctuations changes by a factor of 2. Using this (uncertain) conversion, we estimate that $A_{3D}=\sqrt{P_{3D}4\pi k^3}\sim\delta\rho/\rho$ is in the range 0.02-0.05, which corresponds to a turbulent Mach number $M_{3D}\sim0.1-0.25$ \citep{gaspari13} (see Sect. \ref{sec:icmphysics}). We note that this estimate significantly depends on the 3D geometry of the structure, therefore the value of the Mach number reported here should be considered as indicative.

As a word of caution, we note that in the temperature range of our structure, the emissivity of the plasma in the [0.5-2.0] keV band is not strictly proportional to the emission measure of the plasma. Variations in temperature and metallicity will induce changes in the X-ray emissivity, such that the power spectrum of surface-brightness fluctuations does not depend on underlying density fluctuations alone. For temperatures in the range 1-1.5 keV, the emissivity of the plasma changes by 30\%, which adds another source of uncertainty to the normalization of $A_{3D}$. The slope of the power spectrum is unaffected, however.




\section{Discussion}
\label{sec:disc}

Our \emph{Chandra} observation of the infalling group in A2142 has revealed a wealth of details on the effects of ram-pressure stripping on the gaseous atmosphere of a galaxy group. Here we give some clues on the interpretation of these results. As described in Sect. \ref{sec:tail}, the geometry of the structure is very complex and shows two distinct parts: a straight cometary tail out to $\sim250$ kpc from the core of the group, and an irregular patchy region located beyond (hereafter \emph{flaring wake}). During infall, the gradual increase in the ram pressure ($P_{\rm ram}=\rho_{\rm ICM}v_{\rm rel}^2$) progressively removes shells from the original group atmosphere, which results in a geometry resembling a cone \citep{toniazzo01,heinz03}. The gas located in the outermost regions of the tail thus originates from the outskirts of the group and was stripped first because of its low thermal pressure (see E14). The overall geometry of the tail is reminiscent of a conical 3D geometry whereby the outer regions of the original group atmosphere are located at a larger distance from the core of the group, in agreement with the expectations. We start by discussing the innermost region, then we provide some interpretation of the flaring behavior of the wake. Finally, we discuss the implications of the power spectrum of density fluctuations in the wake. 

\subsection{Straight tail}

As can be seen in Fig. \ref{fig:boxes}, in the inner regions of the tail the surface brightness distribution perpendicular to the group motion is strongly asymmetric. While in the NW direction the emission falls off rapidly, forming a sharp straight edge, in the SE direction the surface-brightness decreases gradually until it reaches the background level. Such a morphology cannot be easily explained by projection effects, as one would expect the sharp edge to extend over the full length of the wake; instead, the width of the wake abruptly increases beyond 250 kpc, which cannot be easily reproduced by projection effects. The sharp edge observed at the NW boundary of the structure suggests that the stripped gas in the innermost regions of the tail is effectively shielded from its environment, for instance, by magnetic draping \citep{asai05,zuhone13}. 

The observed asymmetric morphology of the wake could originate from previous asymmetries in the group atmosphere. Such asymmetries could occur from an AGN outburst in one of the core galaxies, or as a result of the merger activity among the core galaxies. Both processes can lead to sloshing motions of the group atmosphere before or parallel to the stripping.  The same would be true if the group encountered a large turbulent eddy in the cluster outskirts.  Ram pressure stripping removes lower density gas more easily, therefore an initially asymmetric atmosphere can lead to an asymmetric wake.

To illustrate this process, in Fig. 10 we show a series of snapshots from a gas stripping simulation of an elliptical galaxy that is experiencing an AGN outburst (Roediger et al. in prep.). The basic gas stripping simulation is the same as the ``initially extended atmosphere'' run in \citet{roediger15a,roediger15b}; that is, the simulation aims to model the infall of the elliptical galaxy M89 into the Virgo cluster. In addition to the setup described in \citet{roediger15a,roediger15b}, we have added an AGN outburst to the galaxy by injecting $10^{56}$ ergs of thermal energy into two small (1 kpc) spherical volumes over a duration of 5 Myr. These bubbles are overpressured and expand, then rise buoyantly in the galaxy atmosphere over a few 10 Myr. The expanding and rising bubbles push the galactic gas in their path to larger radii where the still-ongoing ICM head wind can strip the gas more easily, giving the galaxy a ``hiccup''. Thus, the AGN outburst leads to a short ($\lesssim100$ Myr) period of enhanced gas stripping that eventually is seen in the wake as a region of higher stripped gas density. In the simulation, the AGN outburst was set up along an axis only a few degrees away from the direction of motion of the galaxy and still introduces some asymmetry in the wake. 

In the group infalling onto A2142, the intense merging and AGN activity observed at the tip of the structure (see Sect. \ref{sec:tip}) is likely to introduce similar disturbances in the gas density distribution of the group, which may then have repercussions on the stripping process. Obviously, this scenario is rather speculative, and tailored numerical simulations are required to test whether core sloshing and/or AGN-inflated bubbles can cause the striking asymmetry shown in Fig. \ref{fig:boxes}.

\begin{figure}
\resizebox{\hsize}{!}{\includegraphics{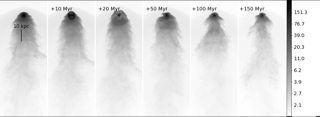}}
\caption{Series of mock X-ray images of a numerical simulation of combined ram pressure stripping and AGN outburst in an elliptical galaxy. The simulation is tuned to reproduce the galaxy M89 \citep{roediger15a,roediger15b} and includes a single 5 Myr long AGN outburst (see text). The time labels show time since the beginning of the outburst. }
\label{fig:sim_elke}
\end{figure}

\subsection{Flaring wake}

At a distance of 250 kpc from the core of the group, the width of the tail rapidly increases and the geometry of the structure no longer resembles a cone, suggesting an abrupt change in the physical conditions. We envisage two possible scenarios to explain this change: \emph{i)} an abrupt ram-pressure stripping event that would have caused a large portion of the gas to be stripped on a short timescale; \emph{ii)} an expansion and mixing of the group gas with the surrounding medium.

In the first case, an abrupt change in the ram pressure may have been caused by the passage of the infalling group through a shock front. Shock fronts are characterized by sharp jumps in both the pressure and density of the ICM, leading to a rapid increase of the surrounding pressure. Merger shock fronts in the ICM typically have Mach numbers in the range 2-3, while more distant and powerful accretion shocks can have Mach numbers of 10 or more \citep{vazza12}. The pressure of the ambient gas going through a Mach 2 shock increases by a factor of $\sim5$, which could be sufficient to strip a large amount of gas on a short timescale. An association between shock fronts and ``jellyfish'' galaxies has been reported by \citet{owers12}, which indicates that the passage through a shock front may indeed cause intense ram-pressure stripping episodes. However, the change in morphology is not associated with an increase in the surface brightness of the tail, which we would have expected in the case where a large amount of gas had been stripped rapidly. The increase in the width of the structure also appears to be much more pronounced toward the NW, which would require a strongly asymmetric geometry of the group prior to infall. 

Alternatively, we also consider the possibility that the stripped gas located beyond 250 kpc from the tip is overpressured compared to the local ICM. In this case, the gas can expand within the surrounding medium and start interacting with the ambient ICM. The gas should therefore start to interact and become gradually heated to the virial temperature of the halo. We would thus expect the gas in this region to be strongly multiphase. Interestingly, this scenario may provide an explanation for the puzzling metallicity map shown in Fig. \ref{fig:spectroimaging}. \citet{buote00} showed that low-resolution X-ray spectra are subject to an ``iron bias'' when fitting the emission of multiphase $\sim1$ keV plasmas with single-temperature models. In this case, the L-shell blend that is used to measure the metallicity is smeared out, resulting in featureless spectra. To investigate this possibility, we used the differential emission-measure model \texttt{c6mekl} \citep{singh96} to fit the spectrum of this region, fixing the metal abundance to the canonical value of 0.25$Z_\odot$. We found that this model provides a good description of the data using a broad emission-measure distribution with temperatures ranging from 1 to 5 keV. Although this solution cannot be statistically favored over a single-temperature model with a low metal abundance, it provides a more realistic description of the processes at work in the outermost regions of the tail. It is therefore ikely that the gas located in the flaring wake beyond 250 kpc from the core of the group is multiphase and is now mixing with the surrounding ICM. 

\subsection{Implications for ICM physics and transport processes}
\label{sec:icmphysics}

E14 showed that the Spitzer conduction timescale in the hot ICM surrounding the structure is on the order of a few Myr. This is at least two orders of magnitude shorter than the survival time of the stripped gas, which implies a strong suppression of thermal conduction in the ICM \citep[see also][]{degrandi16}. However, it was unclear whether conduction was suppressed only at the boundary between the stripped gas and the ICM or if conduction is globally inhibited. The survival of the stripped gas in the flaring wake suggests that transport processes are inhibited even in the region where mixing is taking place, since the size of the stripped tail beyond the breaking point is still several hundred kpc. 

The relatively flat slope of the power spectrum of surface-brightness fluctuations (see Sect. \ref{sec:ps}) provides us with additional evidence for a strong suppression of conductivity in the plasma. As shown in \citet{gaspari14}, Spitzer-like thermal conduction washes out small-scale perturbations in the ICM, thus the slope of the density power spectrum becomes much steeper than the Kolmogorov slope on scales of 100 kpc and less. On the other hand, if conduction is inhibited, Kolmogorov-like velocity power spectra lead to nearly flat density power spectra, in agreement with the observational results provided here. We stress that this result was obtained by considering only the gas within the tail, thus it is independent of any confinement of the gas from its surroundings (e.g., by magnetic draping). Therefore, we conclude that our observations favor a low effective conductivity throughout the bulk of the plasma with a suppression factor $f\lesssim10^{-3}$ as found in the Coma cluster \citep{gaspari13}) and a slow mixing of the merging plasma.

As discussed in Sect. \ref{sec:ps}, in principle, the power spectrum of surface-brightness fluctuations can be used to retrieve an estimate of the level of turbulence in the plasma, since density fluctuations behave as a quasi-passive tracers of velocity fluctuations in the subsonic regime. By using the conversion given in \citet{gaspari13}, $M_{3D}\approx4A_{3D, max}/(L/{\rm 500 kpc})^{0.3}$, with $L$ the injection scale. From the amplitude of fluctuations and given an uncertainty of a factor of 2 in the conversion from $A_{2D}$ to $A_{3D}$ (see Appendix \ref{app:conicalps}), we obtain that the characteristic amplitude of $\delta\rho/\rho$ is in the range $0.02-0.05$ at an injection scale of 250 kpc. According to the above theoretical model, we expect perturbations mainly residing in the isobaric regime and driven by entropy waves $\delta K/K=(5/3)\delta\rho/\rho$ \citep{gaspari14,zhu15}. Temperature fluctuations should be on the same level as the density $\delta T/T=|\delta \rho/\rho|$. Thus, our measurements are consistent with a mild level of turbulence in the stripped gas with a Mach number in the range 0.1-0.25. We note that unfortunately the existing data quality is not sufficient to measure the amplitude of temperature fluctuations as well.

The cascading turbulent motions should lead to a small-scale tangled magnetic field configuration, which may be responsible for suppressing conduction on small scales, in addition to plasma micro-instablities such as mirror and firehose \citep{chandran98,rechester78,ruszkowski10,komarov14}. Chaotic motions can reorganize the magnetic field configuration at the group-ICM interface, which may lead to the breaking of the magnetic drape. This provides a possible explanation for the abrupt change in morphology around the breaking point of the straight tail, where turbulent mixing likely starts to isotropize the stripped gas. 

For an average temperature of $kT=1.5$ keV in the flaring wake and a sound speed $c_s = 625$ km/s, the 3D velocity dispersion should be in the range $\sigma_v = M_{3D}c_s \sim 60-160$ km/s. The turbulent diffusivity is $D_{\rm turb} \equiv L \sigma_v \simeq 10^{31}$ cm$^2$ s$^{-1}$. Thereby, to entirely mix the stripped wake, it would require an additional time $t_{\rm mix} = L^2/D_{\rm turb} \approx 2$ Gyr. We note that for a 1.5 keV plasma with $n_e = 5\times10^{-4}$ cm$^{-3}$ (related to the gas within the wake, not in the hotter ICM) and a suppression factor $f \lesssim 10^{-3}$, the conduction diffusivity is $D_{\rm cond}< 10^{29}$ cm$^2$ s$^{-1}$, that is, a substantially slower transport process than turbulence. Overall, transport processes in the ICM appear highly inhibited and relatively slow, allowing us to observe a wealth of density or temperature perturbations trailing behind infalling structures and later filling the whole cluster atmosphere.

\section{Conclusions}

In this paper, we have reported on deep \emph{Chandra}/ACIS-S observations of the galaxy group falling into A2142, which reveal a very complex morphology and a wealth of phenomena induced by ram-pressure stripping. Our results can be summarized as follows.

We identified the core of the structure as being associated with a compact group of at least three galaxies (G3-G5 in Fig. \ref{fig:tip}), unlike the association that was put forward in E14. The gas in the core of the group has a temperature of $\sim1$ keV, which agrees with the temperature of the gas in the tail. A leading edge can be observed in the direction of the cluster core, demonstrating that the structure is infalling on the main cluster.

The galaxies in the tip of the structure are characterized by a remarkable level of activity. Two of the galaxies are firmly identified as X-ray AGN. The optical spectra of these galaxies indicate clear star formation activity. The radio morphology of the group shows one or several radio galaxies at the tip and a small tail extending in the direction of the X-ray trail. We postulate that all these phenomena are linked to the infall of the group onto the main cluster.

In the region closest to the core of the group, the morphology of the X-ray trail appears straight and asymmetric, with a sharp boundary in the NW direction and a more gradual decline in the opposite direction. The sharp boundary suggests that the stripped gas is efficiently shielded from its environment, for example, by magnetic draping. The observed asymmetry might be induced by a disturbance of the intragroup gas caused by core sloshing and/or AGN outbursts, which could lead to some parts of the group's atmosphere being stripped more efficiently than others. Tailored numerical simulations are required to confirm this scenario.

Beyond $\sim250$ kpc from the core of the group, the tail flares abruptly, reaching a width of $\sim300$ kpc. The morphology of this ``flaring wake'' becomes irregular and patchy, and the sharp boundary observed in the innermost region disappears. The irregular appearance of the tail is reminiscent of turbulent motions and/or mixing with the surrounding ICM. We postulate that the abrupt change in morphology can be due to the passage through a shock front and/or to a breaking of the magnetic drape. 

The temperature structure of the gas in the tail is nearly homogeneous, with a temperature in the range 1-2 keV. The metallicity map of the structure shows a typical metallicity of $\sim0.2Z_\odot$ out to $\sim300$ kpc from the tip and a drop to nearly 0 metallicity beyond this point. This behavior can be explained if the gas in the outermost regions is multiphase and the iron abundance is biased toward low values \citep{buote00}. This observation is consistent with the interpretation that the gas beyond 300 kpc is starting to mix with its environment, resulting in a multiphase configuration.

The power spectrum of surface-brightness fluctuations in the flaring wake exhibits a shallow slope ($P_{2D}\propto k^{-2.3}$) down to the smallest scales accessible to this study ($\sim25$ kpc). The flat slope of the power spectrum and the long survival of small-scale fluctuations in the plasma beyond the ``breaking point'' indicate that thermal conduction is strongly suppressed ($f\lesssim10^{-3}$) within the bulk of the stripped gas, confirming the interpretation put forward in E14. 

The amplitude of the surface-brightness fluctuations is $10-14\%$ depending on the scale. Considering a conical geometry with different viewing angles, our results are consistent with a mild level of turbulence ($M_{3D}\sim0.1-0.25$), which may be responsible for a breaking of the magnetic drape at some distance from the core of the group. The low level of turbulence and suppressed thermal conduction imply a slow ($\sim2$ Gyr) mixing of the infalling gas with the surrounding medium.

\begin{acknowledgements}
We thank Eugene Churazov, Fabrizio Brighenti, and Mauro Roncarelli for useful discussions and the anonymous referee for a constructive review. The scientific results reported in this article are based on observations made by the \emph{Chandra} X-ray Observatory. This research made use of Astropy, a community-developed core Python package for Astronomy \citep{astropy}. Support for this work was also provided by NASA Chandra award number G07-18121X. M.G. is supported by NASA through Einstein Postdoctoral Fellowship Award Number PF5-160137 issued by the Chandra X-ray Observatory Center, which is operated by the SAO for and on behalf of NASA under contract NAS8-03060. Support for this work was also provided by NASA Chandra award number GO7-18121X. Partial support for research for L.R. comes from SAO/NASA grant GO4-15119A to the University of Minnesota.
\end{acknowledgements}

\bibliographystyle{aa}
\bibliography{chandra_a2142}

\appendix
\section{Projected power spectrum in conical geometry}
\label{app:conicalps}

As discussed in Sect. \ref{sec:tail}, the geometry of the stripped gas is complex and overall resembles a conical geometry. The uncertainties in the geometry of the structure can affect how we use the observations to infer the actual shape and slope of the underlying power spectrum.

Here we study how the projected power spectrum depends on the orientation angle with respect to the plane of the sky in a conical geometry. To this aim, we simulated a 3D conical geometry with a gas density profile represented by a beta model leading to an approximately flat projected brightness distribution. The 3D geometry is perturbed by a fluctuation field with a slope of $-8/3$ in Fourier space to match the observed slope (see Sect. \ref{sec:ps}). We then varied the inclination angle of the cone with respect to the plane of the sky, from $\theta=0$ (perfectly aligned with the plane of the sky) to $\theta=\pi/2$ (orthogonal), and projected the resulting images along the line of sight. As shown in Appendix B of \citet{gaspari13}, the underlying beta model introduces power on large scales only and does not affect the slope of the cascade on small scales. We then calculated the power spectrum of the projected images and compared the results with the injected fluctuation field. 

\begin{figure}
\resizebox{\hsize}{!}{\includegraphics{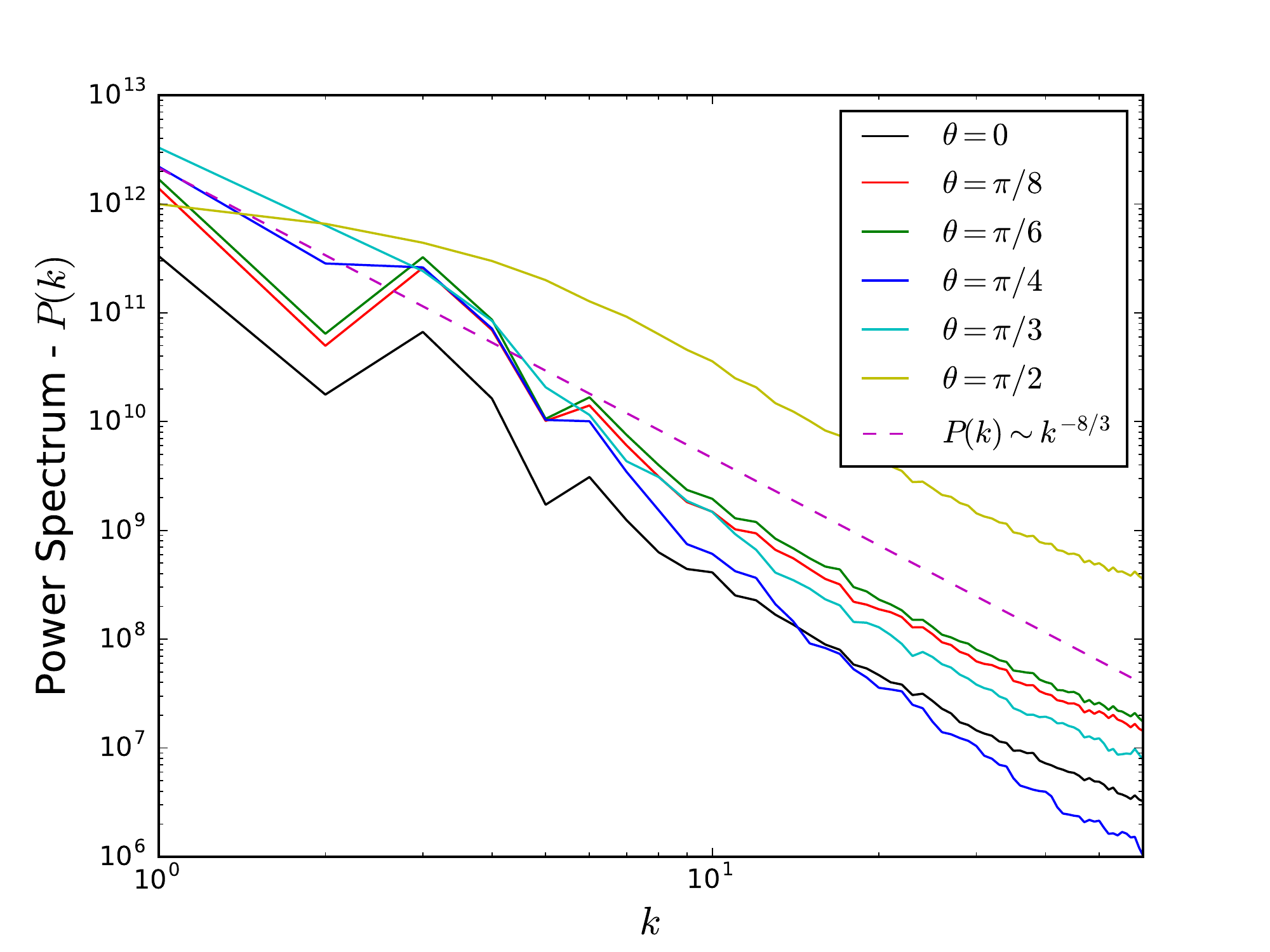}}
\caption{Projected 2D power spectrum (in arbitrary units) for a conical 3D geometry and a perturbation field with a power spectrum $P(k)\sim k^{-8/3}$ (dashed line). The solid curves show the recovered shape of the power spectrum for different inclination angles of the axis of the cone with respect to the plane of the sky. Here $\theta=0$ means that the cone is placed exactly in the plane of the sky, while for $\theta=\pi/2$ is projected along the line of sight.}
\label{fig:coneps}
\end{figure}

In Fig. \ref{fig:coneps} we show the projected power spectra for six different orientation angles compared to the injected slope. We can see that for $\theta=\pi/6$ and lower, the slope of the power spectrum at high $k$ is essentially unaffected by projection effects. However, the normalization of the power spectrum, which is necessary to convert between 2D and 3D power spectra, is in the range $1.2-7.4$, meaning that the uncertainty in the conversion is about a factor $\sim5$. Thus, we conclude that projection leaves the slope almost unchanged, but introduces important systematic uncertainties in the absolute normalization, which is necessary to relate the amplitude of density fluctuations with the turbulent Mach number. This calculation assumes that the 3D geometry of the structure is approximately conical; the conversion factor may vary when the true geometry strongly differs from this assumption.

\end{document}